\documentclass[aps,pra,a4paper, twocolumn, 10pt, showpacs]{revtex4-1} %
\usepackage{bbm, amsmath, amssymb, amsthm, bm, textcomp,graphicx,color}
\usepackage[utf8]{inputenc}
\usepackage[T1]{fontenc}
\usepackage[english]{babel}

\theoremstyle{definition}
\newtheorem{theorem}{Observation}
\newtheorem*{rem}{Remarks}
\newtheorem*{eg}{Examples}

\begin{document}

\title{Tighter quantum uncertainty relations follow from a general probabilistic bound}

\author{Florian Fr\"owis$^{1}$, Roman Schmied$^{2}$, Nicolas Gisin$^{1}$}
\affiliation{$^1$ Group of Applied Physics, University of Geneva, 1211 Geneva, Switzerland\\
$^2$ Department of Physics, University of Basel, Klingelbergstrasse 82, 4056 Basel, Switzerland}
\date{\today}

\begin{abstract}
  Uncertainty relations (URs) such as the Heisenberg-Robertson or the time-energy UR are often considered to be hallmarks of quantum theory. Here, a simple derivation of these URs is presented based on a single classical inequality from estimation theory, a Cram\'er-Rao-like bound. The Heisenberg-Robertson UR is then obtained by using the Born rule and the Schr\"odinger equation. This allows a clear separation of the probabilistic nature of quantum mechanics from the Hilbert space structure and the dynamical law. It also simplifies the interpretation of the bound. In addition, the Heisenberg-Robertson UR is tightened for mixed states by replacing one variance by the quantum Fisher information. Thermal states of Hamiltonians with evenly-gapped energy levels are shown to saturate the tighter bound for natural choices of the operators. This example is further extended to Gaussian states of a harmonic oscillator. For many-qubit systems, we illustrate the interplay between entanglement and the structure of the operators that saturate the UR with spin-squeezed states and Dicke states.
\end{abstract}

\maketitle

\section{Introduction}
\label{sec:introduction}

Uncertainty relations (URs) are tightly connected to quantum mechanics and are often said to be the cornerstone of the theory. For a generic quantum state $\rho$, the product of variances of two noncommuting self-adjoint operators $A,B$ is not vanishing, indicating the impossibility of preparing quantum states with certain properties with respect to all possible observables. Mathematically, this can be expressed by the Heisenberg-Robertson UR \cite{robertson_uncertainty_1929}
\begin{equation}
\label{eq:1}
(\Delta A)^2 _{\rho} ( \Delta B)^2 _{\rho} \geq \frac{1}{4}  \langle i\left [ A,B \right] \rangle_{\rho} ^2,
\end{equation}
with the variance $(\Delta A)_{\rho}^2 = \langle A^2 \rangle_{\rho} - \langle A \rangle_{\rho}^2$ and similarly for $(\Delta B)_{\rho}^2$. In a related spirit, the time-energy UR (in the formulation of Madelstam and Tamm \cite{mandelstam_uncertainty_1945}) connects the variance of the system Hamiltonian $H$ with the time $\Delta t$ it takes to evolve a quantum state $\rho$ to an orthogonal state via
\begin{equation}
\label{eq:2}
(\Delta H)_{\rho} \Delta t \geq \frac{\pi \hbar}{2}.
\end{equation}

Independent of quantum theory, URs also appear in the field of metrology to bound the minimal error on parameter estimates. Among the most famous inequalities is the Cram\'er-Rao bound \cite{Cramer_Mathematical_1945,*RadhakrishnaRao_Information_1945}. Consider the probability distribution that arises from a measurement $A$ and assume that it depends on the value of the parameter $\theta$ (more details are given in Sec.~\ref{sec:derivation}). Holevo \cite{Kholevo_Generalization_1974} derived what he called a generalized Cram\'er-Rao bound, 
\begin{equation}
\label{eq:5}
(\Delta A)^2 F(\theta)  \geq \left(\frac{d}{d\theta}\langle A \rangle\right)^2,
\end{equation}
where $F(\theta)$ is the Fisher information. Later, a quantum version of Eq.~(\ref{eq:5}) was found \cite{Hotta_Quantum_2004,Pezze_Entanglement_2009,Zhong_Optimal_2014}.
One easily continues the list of quantum URs, for example, by mentioning squeezing inequalities \cite{Sorensen_Entanglement_2001,*Sorensen_Many-particle_2001} and bounds on multiparticle entanglement \cite{Pezze_Entanglement_2009,Hyllus_Fisher_2012,*Toth_Multipartite_2012}.

In this paper we discuss the connection between these URs. In particular, we present a proof of (\ref{eq:1}) based on (\ref{eq:5}) by using the Born rule and the Schr\"odinger equation. Hence, this derivation provides insight into the influences of different aspects of quantum theory on the UR, that is, its probabilistic nature, the Hilbert space structure and the dynamical law. In addition, it allows a different view of the Heisenberg-Robertson UR. The interpretation of the inequality as the mathematical expression of Heisenberg's microscope argument \cite{heisenberg_uber_1927} is hard to maintain (which is in line with previous contributions \cite{hilgevoord_uncertainty_1991,*ozawa_uncertainty_2004,*busch_proof_2013}). Note that, as discussed later, the Heisenberg-Robertson UR can also be seen as a looser version of the Schr\"odinger UR \cite{Schrodinger_Heisenbergschen_1930}, which is incompatible with (\ref{eq:5}).

In the second part of the paper, we focus on a tighter version of the Heisenberg-Robertson UR for mixed states (see also \cite{Hotta_Quantum_2004,Pezze_Entanglement_2009,Zhong_Optimal_2014}).
There the quantum Fisher information (QFI) $\mathcal{F}$ appears and replaces one variance in Eq.~(\ref{eq:1}):
\begin{equation}
\label{eq:3}
( \Delta A)^2_{\rho} \mathcal{F}_{\rho}(B) \geq \langle i [A,B] \rangle_{\rho}^2.
\end{equation}
The variance is always greater than or equal to a quarter of the QFI, where equality holds for pure states. The QFI, which is a convex function, is used in quantum metrology to quantify how well different values of a (partially) unknown parameter can be distinguished \cite{helstrom_quantum_1976,holevo_probabilistic_2011,Braunstein_Statistical_1994}. More recently, its role in multi-particle entanglement was discovered \cite{Pezze_Entanglement_2009,Hyllus_Fisher_2012,Toth_Multipartite_2012}. 
In particular, we investigate under which circumstances Eq.~(\ref{eq:3}) is tight (compare to \cite{Braunstein_Statistical_1994,Helstrom_Estimation_1974,*Braunstein_Generalized_1996,Hotta_Quantum_2004,Zhong_Optimal_2014}). One can show that for every pair $(\rho,B)$ there exists an optimal $A$ such that one finds equality. Here, we present a whole class of such instances for thermal states of evenly gapped Hamiltonians and $A,B$ as linear combinations of the corresponding ladder operators. It turns out that this can even be generalized for Gaussian states of the harmonic oscillator, given more specific choices of $A,B$. In addition, we discuss many-qubit systems with highly entangled states $\rho$ and local operators $B$. The entanglement within a reduced density operators of a small subset of  qubits significantly influences the structure of the optimal $A$. We illustrate this by presenting the optimal $A$ for so-called ``over-squeezed'' spin-squeezed states \cite{Kitagawa_Squeezed_1993} and the Dicke state \cite{Dicke_Coherence_1954}.

\section{Heisenberg-Robertson UR from a classical bound}

In this section, we first give a simple proof of (\ref{eq:5}) and connect it then to (\ref{eq:1}). Later, we discuss the connection to other URs such as (\ref{eq:2}) and draw some  conclusions from the presented derivation.

\subsection{Derivation}
\label{sec:derivation}

The following derivation of Eqs.~(\ref{eq:1}) and (\ref{eq:3}) is adaptable to continuous probability distributions. For the sake of simplicity, however, we focus on the discrete case. Consider a metric space of probability distributions $\left\{ p_i \right\}_{i=1}^d$ for $d$ discrete events $i$. In addition, one assigns measurement outcomes $a_i$ to each $i$. The expectation value and the variance of this observable read $\langle A \rangle = \sum_i a_i p_i$ and $(\Delta A)^2 = \sum_i (a_i - \langle A \rangle_{})^2 p_i$, respectively.

Suppose that one introduces a differentiable curve through the space of probability distributions; parametrized by a real variable $\theta $ from an open interval in $\mathbbm{R}$. Hence, points on this line depend on $\theta$, $p_i = p_i(\theta)$, and we only consider points along this curve in the following. One defines the Fisher information as 
\begin{equation}
\label{eq:4}
F(\theta) = 4\sum_i \left(  \frac{d}{d\theta} \sqrt{p_i} \right)^2.
\end{equation}
Every point $\left\{ p_i(\theta) \right\}_i$ carries certain information about $\theta$. The Fisher information is a way to quantify how distinguishable probability distributions with similar $\theta$ are.

Consider $(\Delta A)^2$ and $F(\theta)$ as squared norms of vectors with entries $(a_i - \langle A \rangle_{}) \sqrt{p_i}$ and $2 \frac{d}{d\theta} \sqrt{p_i}$, respectively. Thus, by a single application of the Cauchy-Schwarz inequality $\lVert x \rVert^2 \lVert y \rVert^2 \geq \left| \left\langle x | y \right\rangle  \right|^2$, one finds that 
\begin{equation}
\label{eq:5a}
\begin{split}
  (\Delta A)^2 F(\theta) \geq& \left[ \sum_i (a_i -  \langle A \rangle_{}) \sqrt{p_i}\,  2 \left( \frac{d}{d\theta} \sqrt{p_i} \right)\right]^2  \\ =&\left( \sum_i a_i\frac{d}{d \theta}p_i - \langle A \rangle_{} \frac{d}{d\theta }\sum_i p_i\right)^2.
\end{split}
\end{equation}
Since $\frac{d}{d\theta} \sum_i p_i = 0$, one ends up with Eq.~(\ref{eq:5}).
For an alternative proof of Eq.~(\ref{eq:5}), see Ref.~\cite{Kholevo_Generalization_1974}.

The quantum formalism is now applied to inequality (\ref{eq:5}). Instead of dealing with general transformations and measurements, we limit ourselves to unitary evolution and projective measurements. The following operators thus act on the Hilbert space $\mathbbm{C}^D$ with $d \leq D \in \mathbbm{N}$. In quantum mechanics, one has a density operator $\rho$ and a complete set of orthogonal projectors $\Pi_i$ associated with the events $i$ such that the probabilities are calculated via the Born rule, that is, $p_i = \mathrm{Tr}\rho \Pi_i$. Next, assume that the parametrization in the space of probability distributions is caused by a unitary transformation governed by the Schr\"odinger equation. In the Heisenberg picture, the operator  $A = \sum_i a_i \Pi_i$ transforms via
\begin{equation}
\label{eq:6}
\frac{d}{d \theta} A = -i [A,B],
\end{equation}
where $B$ is a self-adjoint operator that generates the evolution. Equivalently, the time dependence of the state in the Schr\"odinger picture reads $\rho(\theta) = \exp(-i B \theta) \rho_0 \exp(i B \theta)$.
As the last step, note that \cite{Braunstein_Statistical_1994}
\begin{equation}
\label{eq:8}
F(\theta) \leq \mathcal{F} \mathrel{\mathop:}= \max_{\left\{ \Pi_i \right\}_i} F(\theta) \leq 4( \Delta B)^2_{\rho},
\end{equation}
where the maximization is over all possible measurement settings $\left\{ \Pi_i \right\}_i$ while keeping the state $\rho$ and the dynamics (\ref{eq:6}) fixed. (It is sufficient to restrict ourselves to von Neumann measurements \cite{Braunstein_Statistical_1994}.) The last inequality in Eq.~(\ref{eq:8}) is a strict equality for pure states. The maximal Fisher information $\mathcal{F}$ is the QFI, which is a convex function in $\rho$. Interestingly, it turns out that the quantum Fisher metric (which is the basis for the QFI) and the Bures metric are identical up to a factor 4 \cite{Braunstein_Statistical_1994}, which underlines the importance of these metrics. For unitary transformations and given the spectral decomposition  $\rho(\theta) = \sum_i q_i \left| \psi_i(\theta) \right\rangle\!\left\langle \psi_i(\theta)\right| $, it reads \cite{Braunstein_Statistical_1994}
\begin{equation}
\label{eq:9}
\mathcal{F} =2 \sum_{i,j} \frac{(q_i-q_j)^2}{q_i+q_j} \left| \left\langle \psi_i (\theta)\right| B \left| \psi_j (\theta) \right\rangle \right|^2.
\end{equation}
Since $\left| \psi_i(\theta) \right\rangle  = \exp(-i B \theta) \left| \psi_i(0) \right\rangle $, Eq.~(\ref{eq:9}) is independent of $\theta$ and the QFI is denoted by $\mathcal{F} \equiv \mathcal{F}_{\rho}(B)$.
Equations~(\ref{eq:5}) and (\ref{eq:6}) and the first inequality in (\ref{eq:8}) directly lead to Eq.~(\ref{eq:3}); using the second inequality in Eq.~(\ref{eq:8}) leads to the Heisenberg-Robertson UR (\ref{eq:1}). The asymmetry of Eq.~(\ref{eq:3}), where $B$ generates the evolution and $A$ is the measurement operator, is lifted in Eq.~(\ref{eq:1}). This is because, for pure states, it has a double role. On the one hand, the variance defines the infinitesimal line element in the evolution of the state and, on the other hand, it is part of the measurement uncertainty. Note that in Eq.~(\ref{eq:3}) replacing $(\Delta A)^2_{\rho}$ by $\mathcal{F}_{\rho}(A)/4$ is in general not possible.

\subsection{Connections between different URs}
\label{sec:conn-betw-diff}

The primitive inequality (\ref{eq:5}) and its specialized quantum version (\ref{eq:3}) not only lead to the Heisenberg-Robertson UR, but also give rise to several quantum URs.

First of all, the presented derivation establishes a strong link between the Heisenberg-Robertson UR and the quantum Cram\'er-Rao bound, which complements previous discussions on this topic \cite{Helstrom_Estimation_1974,Braunstein_Generalized_1996}. Next the connection between the Heisenberg-Robertson UR and the time-energy UR was already implicitly shown in Ref.~\cite{mandelstam_uncertainty_1945}. 
Note that, by starting with the tighter bound (\ref{eq:3}), one arrives at a tighter time-energy UR, where the variance of the Hamiltonian is replaced by the QFI \cite{frowis_kind_2012}. Finally, spin-squeezed inequalities \cite{Sorensen_Many-particle_2001,Sorensen_Entanglement_2001,Wang_Spin_2003}, which define spin-squeezing and give sufficient conditions for entanglement in composed spin systems, are also direct consequences of Eq.~(\ref{eq:3}).

\subsection{Interpretation of the Heisenberg-Robertson UR}
For Robertson, the primary motivation to prove Eq.~(\ref{eq:1}) was to find a mathematical formulation of Heisenberg's microscope argument \cite{heisenberg_uber_1927,robertson_uncertainty_1929}: Assume that the position of an electron with a ``well-determined'' momentum is measured by a light microscope. The precision of this measurement depends on the wave length of the photons that scatter with the electron. A large energy of the photons results  in a large momentum kick of the electron. Hence, the smaller the uncertainty $\delta x$ of the position estimation, the larger the uncertainty $\delta p$ of the momentum of the electron afterward. Then Heisenberg heuristically showed that $ \delta x \, \delta p \approx h$.

It is repeatedly argued that the variance of an operator is a poor figure of merit to quantify the disturbance of a state by the measurement and that Eq.~(\ref{eq:1}) does not properly reflect Heisenberg's argument. In papers such as \cite{hilgevoord_uncertainty_1991,ozawa_uncertainty_2004,busch_proof_2013}, more sensible mathematical formulations of measurement-induced disturbances were developed and similar inequalities were formulated. However, what is then a correct interpretation of Eq.~(\ref{eq:1})? The derivation of the Heisenberg-Robertson UR presented in this paper offers the following solution. With the identification of $A = \hat{x}$ and $B = \hat{p}$ as position and momentum operators, it becomes evident through Eq.~(\ref{eq:3}) that the momentum operator generates a shift $\exp(-i x \hat{p})$ of the electron, that is, it prepares the state before measurement. In particular, it does not reflect the uncertainty in momentum after the measurement. Hence, one way to see the Heisenberg-Robertson UR here is that it is a special instance of the Cram\'er-Rao bound, since the uncertainty in the position measurement expressed as $ (\Delta \hat{x})^2 / (\frac{d}{dx}\langle \hat{x} \rangle)^{2}$ can be bounded from below by $1/\mathcal{F}_{\rho}(\hat{p})$.

Note, however, that the Heisenberg-Robertson UR is also a consequence of the Schr\"odinger UR 
\begin{equation}
\label{eq:7}
(\Delta A)^2 _{\rho} ( \Delta B)^2 _{\rho} \geq \frac{1}{4}  \langle i\left [ A,B \right] \rangle_{\rho} ^2 + \frac{1}{4} \langle \left\{ \bar{A},\bar{B} \right\} \rangle_{\rho}^2,
\end{equation}
with $\bar{X} = X-\langle X \rangle_{\rho}$. Equation (\ref{eq:1}) trivially results from relaxing Eq.~(\ref{eq:7}) by dropping the last (positive) term. With basic two-level examples, one can show that Eqs.~(\ref{eq:3}) and (\ref{eq:7}) are incompatible, meaning that one can not be derived from the other \cite{Toth_Unpublished_2015}.

\section{States that saturate the tighter bound}
It is interesting to study cases in which quantum URs are saturated. For generic mixed states it holds that $\mathcal{F}_{\rho}(B) < 4( \Delta B)^2_{\rho}$. On the other hand, one can easily see that for any pair $(\rho,B)$, there exists an optimal operator $A_{\mathrm{opt}}$ such that Eq.~(\ref{eq:3}) is tight. First, note that the choice $a_i = c \dot{p_i}/p_i + \langle A \rangle$, for any constant $c \in \mathbbm{R}$, parallelizes the vectors $\left\{2  \frac{d}{d\theta} \sqrt{p_i}  \right\}_i$ and $\left\{ (a_i-\langle A \rangle)\sqrt{p_i} \right\}_i$ used in Eq.~(\ref{eq:5a}). Then, one finds equality in the Cauchy-Schwarz inequality and hence $(\Delta A)^2 F(\theta) = (\frac{d}{d\theta} \langle A \rangle)^2$. Second, the right choice of the measurement basis $\left\{ \Pi_i \right\}_i$ leads to equality in the first part of Eq.~(\ref{eq:8}). Hence, one can always find  $A_{\mathrm{opt}}$. However, this includes the diagonalization of the so-called symmetric logarithmic derivative \cite{Braunstein_Statistical_1994} and in general does not lead to clear expressions.

In this section, we first show that thermal states of Hamiltonians with an evenly gapped spectrum saturate the tighter bound (\ref{eq:3}) if $A,B$ are certain linear combinations of the corresponding ladder operators. Second, we study many-qubit systems and illustrate the influence of so-called bipartite entanglement on the structure of $A_{\mathrm{opt}}$.

\subsection{Thermal states and ladder operators}
\label{sec:therm-stat-ladd}

\begin{theorem}\label{obs1}
  Consider a Hamiltonian $H$ with spectral decomposition $H =
  \sum_{m=0}^{M} m \sum_{\alpha = 1}^{\Delta_m} \left| m,\alpha
  \right\rangle\!\left\langle m,\alpha\right| $, where $\alpha$ labels the $\Delta_m$-fold degeneracy. There
  exist ladder operators $L^{+}$ and $L^{-} = L^{+ \dagger}$ with
  $L^{\pm} \left| m,\alpha \right\rangle = c_{m,\alpha}^{\pm} \left|
    m\pm 1,\alpha \right\rangle $, where $c_{m,\alpha}^{\pm} \in
  \mathbbm{R}$ and $c_{0,\alpha}^{-} = c_{M,\alpha}^{+} = 0$. Then
the Gibbs state $\rho = \exp(- \beta H)/Z$ [with the inverse temperature $\beta \geq 0$ and the normalization $Z = \mathrm{Tr} \exp(-\beta H)$] saturates inequality (\ref{eq:3}) for the choices
  $A = L^{+} + L^{-}$ and $B = i(L^{+} - L^{-})$ .
\end{theorem}
The proof is a straightforward calculation and is presented in Appendix \ref{sec:proof-observ-refobs1}.

\begin{rem}
  First, this statement holds even in the limit $M\rightarrow
  \infty$. Second, with $(\rho,A,B)$ from Observation \ref{obs1}, one
  can show that the triple $(\rho^{\otimes n}, \sum_{i=1}^n A^{(i)},
  \sum_{i=1}^n B^{(i)})$ also saturates Eq.~(\ref{eq:3}), where $n \in
  \mathbbm{N}$, $A^{(i)} \equiv \mathit{id}^{\otimes i-1} \otimes A
  \otimes \mathit{id}^{\otimes n- i}$ and $\mathit{id}$ the identity
  operator on a single system.
\end{rem}

\begin{eg}
  (1) Any rank-two operator $\rho = g \left| \psi_0
  \right\rangle\!\left\langle \psi_0\right| + (1-g) \left| \psi_1
  \right\rangle\!\left\langle \psi_1\right| $ with $\left\langle
    \psi_i | \psi_j \right\rangle = \delta_{ij}$ and $g \in [1/2,1]$
  is a thermal state of the Hamiltonian $H = \left| \psi_0
  \right\rangle\!\left\langle \psi_0\right|$. The corresponding ladder
  operators read $L^{+} = \left| \psi_1 \right\rangle\!\left\langle
    \psi_0\right| $ and $L^{-} = L^{+ \dagger}$. Therefore,
  Observation \ref{obs1} applies.

  (2) With the second remark, one can extend the first example to $n$
  two-level systems (qubits). In the following, an operator of the
  form $\sum_{i=1}^n A^{(i)}$ is called local. Instances are the
  angular momentum operators $J_k = \frac{1}{2} \sum_{i=1}^n \sigma_k^{(i)}$,
  with the Pauli operators $\sigma_k$ for $k \in \left\{ x,y,z
\right\}$. A tensor product of thermal states, say, in the $x$
  basis, $\rho = \left( g\left| + \right\rangle\!\left\langle +\right|
    + (1-g) \left| - \right\rangle\!\left\langle -\right| \right)
  ^{\otimes n}$, can be seen as a spin-coherent state that is
  polarized in the $x$ direction and that was subject to local phase noise
  \cite{nielsen_quantum_2010}. This state is a classical resource
  in frequency estimation \cite{huelga_improvement_1997}. Since bound
  (\ref{eq:3}) is tight for the choices $A = J_y$ and $B = J_z$, one
  directly calculates the QFI to be $\mathcal{F}_{\rho}(J_z) =
  (2g-1)^2 n$, which corresponds to the so-called standard quantum limit for
  partially dephased spin-coherent states.

  (3) The thermal state of $J_z$, $\rho$, with the choices $A = J_x$
and $B = J_y$ leads to $ (\Delta A)^2_{\rho} = n/4$, $\mathcal{F}_{\rho}(B) =
\tanh^2(\beta/2) n$ and
  $ \langle i [A,B] \rangle_{\rho} = - \langle J_z \rangle_{\rho} = \frac{1}{2} n \tanh (\beta/2)$.

  (4) For single-mode photonic systems, the thermal states with
  respect to the number operator $a^{\dagger} a$ saturate the bound
  (\ref{eq:3}) with the quadrature operators $A \equiv \hat{x}= (a + a^{\dagger})/\sqrt{2}$ and $B \equiv \hat{p}= i(a - a^{\dagger})/\sqrt{2}$, where $i[A,B] = -\mathit{id}$. One finds $\langle \Delta A^2 \rangle_{\rho} = \frac{1}{2}\coth(\beta/2)$ and $\mathcal{F}_{\rho}(B) =
2\tanh(\beta/2) $, which when multiplied are equal to one.
\end{eg}

For the last system, one can generalize the example even further. Let us consider the important class of Gaussian states, 
\begin{equation}
\label{eq:25}
\rho_{\mathrm{G}} = D_{\alpha}S_{\xi} \rho_{\mathrm{th}} S_{\xi}^{\dagger} D_{\alpha}^{\dagger}.
\end{equation}
Here $\rho_{\mathrm{th}}$ is a single-mode photonic state as in example (4), $S_{\xi} = \exp[\frac{1}{2}r(e^{i \theta} a + e^{-i \theta} a^{\dagger})]$ with $\xi = r e^{i \theta}$ is the squeezing operation, and $D_{\alpha} = \exp(\alpha a^{\dagger} + \alpha^{*} a)$ is the displacement operation. By direct calculation, one can show that for the choices $A = (e^{-i\theta/2} a + e^{i \theta} a^{\dagger})/\sqrt{2}$ and $B = i( e^{-i\theta/2} a -  e^{i \theta} a^{\dagger})/\sqrt{2}$, the triple $(\rho_{\mathrm{G}},A,B)$ saturates Eq.~(\ref{eq:3}). Compared to the previous example, we now find $\langle \Delta A^2 \rangle_{\rho} = \frac{1}{2}\exp(-2r)\coth(\beta/2)$ and $\mathcal{F}_{\rho}(B) =2\exp(2r)\tanh(\beta/2) $, that is, the variance and QFI become squeezed and antisqueezed, respectively.

\subsection{Entanglement influences the structure of the optimal operators}
\label{sec:entangl-infl-struct}

This section is dedicated to a specific example, namely, pure many-qubit states and local operators for $B$ in Eq.~(\ref{eq:3}). With this choice for $B = \sum_i B_i^{(i)}$, Eq.~(\ref{eq:3}) potentially detects multipartite entanglement. Let us fix the operator norm of all addends $\lVert  B_i^{(i)}\rVert = 1/2$. If $\mathcal{F}_{\rho}(B) > n$, then $\rho$ is entangled \cite{Pezze_Entanglement_2009}. Furthermore, the larger $\mathcal{F}_{\rho}(B)$ is, the larger the so-called entanglement depth. In a rough approximation, if $\mathcal{F}_{\rho}(B) \gtrsim k n$ for an integer $k \in \left\{ 1,\dots,n \right\}$, then $\rho$ must contain at least $k$-partite entanglement (see Ref.~\cite{Toth_Multipartite_2012,Hyllus_Fisher_2012} for the exact statements). 

Here we discuss the connection between entanglement within reduced density operators and the structure of the optimal $A$ in Eq.~(\ref{eq:3}). In particular, we 
illustrate this for two classes of states:  (one-axis twisted) spin-squeezed states \cite{Kitagawa_Squeezed_1993} and Dicke states \cite{Dicke_Coherence_1954}. Spin-squeezed states generated by so-called one-axis twisting is a class of states introduced in \cite{Kitagawa_Squeezed_1993} (see also \cite{Haake_Quantum_2010}, where similar states appear in  kicked top models). They are defined as
\begin{equation}
\label{eq:14}
\left| \mathrm{S}(\mu) \right\rangle = \exp(-i \nu J_x) \exp(-i \frac{1}{2} \mu J_z^2) \left| + \right\rangle ^{\otimes n},
\end{equation}
where $| + \rangle $ is an eigenstate of $\sigma_x$, $\mu$ is the squeezing strength and $\nu$ parametrizes a local rotation such that the maximal variance is along the $y$ axis. (The value of $\nu$ only depends on $n$ and $\mu$ and is omitted in the following.) Dicke states $|m \rangle $ are defined as $n$-qubit symmetric states that are eigenstates of $J_z$ with eigenvalue $m$, where $m \in \left\{ -n/2, \dots,n/2 \right\}$.

Both states exhibit large variances with respect to $J_y$, depending on the values of $\mu$ and $m$, respectively. The exact values for $| \mathrm{S}(\mu) \rangle $ are presented in \cite{Kitagawa_Squeezed_1993}. Qualitatively, for large $n$, $(\Delta J_y)^2$ rapidly increases from $n/4$ to roughly $n^2/8$ by changing $\mu$ from zero to $O(n^{-1/2})$. Dicke states are genuine multipartite entangled (apart from $m = -n/2,n/2$) and exhibit a large variance with respect to $J_y$, $(\Delta J_y)_{\left| m \right\rangle }^2 = n^2/8 + n/4 -  m^2/2$. For values of $m = O(1)$, one finds a quadratic scaling in the variance. In summary, both state families can show a large entanglement depth.

We now turn to the question of which operators $A$ can optimally bound this large QFI via the inequality 
\begin{equation}
\label{eq:15}
\frac{1}{n}\mathcal{F}_{\rho}(J_y) \geq \frac{\langle i [A,J_y] \rangle_{\rho}^2}{n(\Delta A)^2_{\rho}}. 
\end{equation}
In essence, choosing a local $A$ leads to the well-known spin-squeezing inequalities \cite{Sorensen_Many-particle_2001,Sorensen_Entanglement_2001,Wang_Spin_2003}. For small squeezing strength up to $\mu = O(n^{-2/3})$, a local $A$ is close to optimal for $| \mathrm{S}(\mu) \rangle $, but for larger $\mu$, a local $A$ is not sufficient \cite{Pezze_Entanglement_2009}. In contrast, Dicke states are not spin squeezed at all and no local $A$ gives a bound that witnesses entanglement \cite{Wang_Spin_2003}. To find better bounds for oversqueezed and Dicke states, we therefore consider polynomials of local operators. We define 
\begin{equation}
\label{eq:20}
A_k = \sum_{l_1,\dots,l_k \in \left\{ x,y,z \right\}} c_{l_1,\dots,l_k}J_{l_1}\dots J_{l_k}
\end{equation}
as $k$th-order polynomial in the collective operators $J_l$. The tensor $c$ is chosen such that $A_k$ is self-adjoint.

\begin{figure}[htbp]
\centerline{\includegraphics[width=\columnwidth]{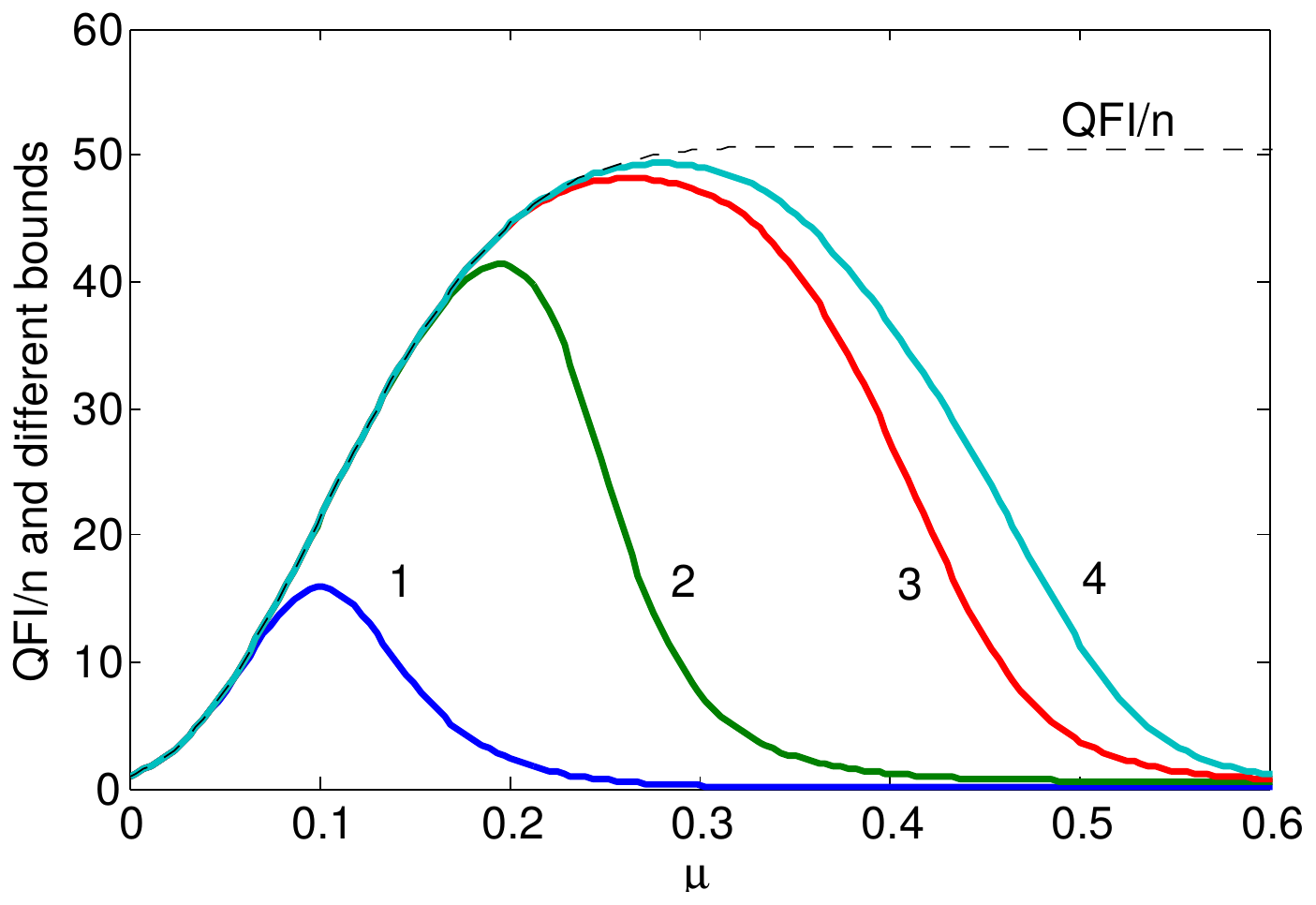}}
\caption[]{\label{fig:SSSOptimalA} (Color online) Comparison of the left- (dashed) and right- (solid) hand sides in Eq.~(\ref{eq:15}) for $| \mathrm{S}(\mu) \rangle $ with $n = 100$. The numbers next to solid lines indicate the $k$ used for a numerical search within the ansatz set (\ref{eq:20}) to maximize the right hand side of Eq.~(\ref{eq:15}). Choosing $k = 1$ corresponds to the spin-squeezed inequalities. Clearly shown is the limited range of each $A_k$ for a tight bounding of the QFI. We remark that the presented curves are lower bounds on the actual optimal $A_k$.}
\end{figure}

For $| \mathrm{S}(\mu) \rangle $ we numerically determined the optimal $A_k$ for small $k$ (see Fig.~\ref{fig:SSSOptimalA} for an example). We find for several instances with up to $n=1000$ that one can increase the range of $\mu$ where the bound (\ref{eq:15}) is tight by increasing $k$. However, an operator $A_k$ with small $k$ that saturates  (\ref{eq:15}) for all $\mu$ does not seem to exist.

The result of the optimal $A$ for Dicke states is simpler. The operator $A_2  =c \left\{  J_x,J_z\right\} + (1-c) J_z$ with $c = 1/(1+2|m|)$ leads to equality in Eq.~(\ref{eq:15}), that is, a quadratic operator is optimal.

What is the reason for the different results for these two state classes? A closer look to the entanglement structure of the reduced density matrices gives some hints. First of all, persistent entanglement after tracing out qubits is not a necessary condition for large variance of local operators. It suffices to have classical correlations between (almost) all pairs of qubits. However, it is necessary for a reduced variance of $A$ \cite{Pezze_Entanglement_2009,Wang_Pairwise_2002,*Toth_Spin_2009}. Both one-axis twisted spin-squeezed states and Dicke states exhibit entanglement within reduced density operators. However, as explicitly shown in Appendix \ref{sec:reduc-dens-matic}, the bipartite entanglement in $| \mathrm{S}(\mu) \rangle $ decays exponentially in $n$, while it only decays algebraically (between $1/n^2$ and $1/n$) for Dicke states.

For spin-squeezed states with increasing $\mu$, we have to continuously increase $k$ to benefit from the entanglement in larger blocks of qubits, which is more persistent than in smaller groups (see Appendix \ref{sec:reduc-dens-matic}). For Dicke states, this is not necessary. The reason why we need there quadratic instead of linear operators is the second condition for a good bound on the QFI: Dicke states are not polarized enough to have a large enough numerator on the left-hand side of Eq.~(\ref{eq:15}). In contrast, the polarization for quadratic operators is sufficient.

\section{Summary and outlook}
In this paper, quantum URs have been connected to a simple inequality from estimation theory. Starting with the inequality (\ref{eq:5}), one goes via Eq.~(\ref{eq:3}) to the Heisenberg-Robertson UR (\ref{eq:1}). With special choices for the operators in Eq.~(\ref{eq:1}), one ends up with the time-energy UR (\ref{eq:2}). Equation (\ref{eq:3}), which is a tighter version of the Heisenberg-Robertson UR, is the basis for other well-known and useful inequalities. In particular, if operator $B$ in Eq.~(\ref{eq:3}) is restricted to local operators in many-qubit systems, one can use Eq.~(\ref{eq:3}) as a simple and efficient bound on multi-particle entanglement. All together, these connections contribute to a broader picture of the structure of quantum mechanics in terms of URs.

The presented derivation provides a clear view on the structure of the Heisenberg-Robertson UR by separating the probabilistic nature of quantum mechanics from the dynamical law and the Hilbert space formalism. Thus, it may help in developing an axiomatic approach of quantum mechanics based on physical principles.
In addition, the classical primitive (\ref{eq:5}) can be used to investigate alternative probabilistic theories. For example, one can keep the Born rule (and the Hilbert space structure) and alter the dynamical law. In this context, the study of collapse models \cite{bassi_models_2013} could be of interest. Collapse models are variations of quantum mechanics. The Schr\"odinger equation is modified to enforce the collapse of spread wave functions of massive objects to localized packages without physical measurement. The altered dynamical law with non-unitary character may give rise to a different Heisenberg-Robertson bound. This could lead to different predictions and therefore additional experimental possibilities to falsify one theory or the other.

\textit{Acknowledgments.--- } 
We thank Otfried G\"uhne, Gebhard Gr\"ubl and Wolfgang D\"ur for critical remarks on an early version of the manuscript. F.F.~acknowledges the hospitality of the Centro de Ciencias de Benasque Pedro Pascual (Spain), where part of this work was advanced. This work was supported by the Austrian Science Fund (FWF), Grant No.~J3462, and by the COST Action No.~MP1006.
\appendix
\begin{widetext}
  \section{Proof of observation \ref{obs1}}\label{sec:proof-observ-refobs1}
  \begin{proof}[\nopunct] The spectral decomposition of the Gibbs
    state reads $\rho = \sum_m q_m \sum_{\alpha} \left| m,\alpha
    \right\rangle\!\left\langle m,\alpha\right|$ with $q_m = e^{-\beta
      m}/Z$. With Eq.~(\ref{eq:9}) one finds
    \begin{equation}
      \label{eq:10}
      \mathcal{F}_{\rho}(B) = \sum_{m<n} \frac{4(q_m-q_n)^2}{q_m+q_n} \sum_{\alpha,\alpha^{\prime}}\left| \left\langle m,\alpha \right| B \left| n,\alpha^{\prime} \right\rangle  \right|^2\!\!.
    \end{equation}
    It holds that $\left| \left\langle m,\alpha \right| B \left|
        n,\alpha^{\prime} \right\rangle \right|^2 = c_{m,\alpha}^{+ 2}
    \delta_{n,m+1} \delta_{\alpha,\alpha^{\prime}}$. We abbreviate
    $C_m^{\pm} = \sum_{\alpha} c_{m,\alpha}^{\pm 2}$. Since
    $(q_m-q_{m+1})^2/(q_m+q_{m+1}) = q_m
    (1-e^{-\beta})^2/(1+e^{-\beta})$, one finds
    \begin{equation}
      \label{eq:12}
      \mathcal{F}_{\rho}(B) = 4 \frac{(1-e^{-\beta})^2}{1+e^{-\beta}}\sum_{m} q_m  C_m^{+}.
    \end{equation}
    Note that since $L^{- \dagger} = L^{+}$, one has $c_{m,\alpha}^{+} =
    c_{m+1,\alpha}^{-}$. Then, with arguments similar to those for the QFI,
    one finds that
    \begin{equation}
      \label{eq:11}
      \begin{split}        
      (\Delta A)^2_{\rho} &= \langle L^{+}L^{-} \rangle_{\rho} +
      \langle L^{-}L^{+} \rangle_{\rho} = 
      \sum_m q_m (C^{-}_m +
      C^{+}_m) = \sum_m q_m (C^{+}_{m-1} + C^{+}_m) = 
      \sum_m
        (q_{m+1} + q_m)C^{+}_m \\ & =(1+e^{-\beta})\sum_m q_mC_m^{+},
      \end{split}
    \end{equation}
    where the second to last equality is only due to a reindexing of the
    first part of the sum. Now, one notices that, up to a constant,
    Eqs.~(\ref{eq:12}) and (\ref{eq:11}) sum over the same addends and
    thus, using Cauchy-Schwarz inequality for parallel vectors, it holds that
    \begin{equation}
      \label{eq:13}
      \mathcal{F}_{\rho}(B)  (\Delta A)^2_{\rho} =4 (1-e^{-\beta})^2 \left(\sum_{m} q_m  C_m^{+}  \right)^2.
    \end{equation}
    The right-hand side of Eq.~(\ref{eq:13}), however, is equivalent to $\langle i
    [A,B] \rangle_{\rho}^2 = \langle 2[L^{+},L^{-}] \rangle_{\rho}^2$
    since
    \begin{equation}
      \label{eq:16}
      \langle [L^{+},L^{-}] \rangle_{\rho} = \sum_m q_m (C_m^{-} -
      C_m^{+}) = 
      \sum_m q_m (C_{m-1}^{+} - C_m^{+}) = \sum_m
      (q_{m+1} - q_m) C_m^{+} 
      = (1-e^{-\beta})\sum_m q_m
        C_m^{+}.\qedhere
    \end{equation}
  \end{proof}

  \section{Reduced density matices of spin-squeezed and Dicke states}
  \label{sec:reduc-dens-matic}

  Here, we present some calculations to determine the amount of
  entanglement in reduced states of symmetric many-qubit systems.

  \subsection{General Formalism} For symmetric states, it is most
  convenient to work in the Dicke basis. Dicke states $| m \rangle =
  \mathrm{sym}(\left| 0 \right\rangle ^{\otimes m+n/2} \otimes \left|
    1 \right\rangle ^{\otimes m-n/2})$ are symmetric eigenstates of
  $J_z$ with eigenvalue $m \in \left\{  -n/2,\dots,n/2\right\}$. For simpler
  expressions, we switch to a different notation of Dicke states and
  write $ | n,k \rangle = \mathrm{sym} (\left| 0 \right\rangle ^{\otimes n-k} \otimes \left| 1 \right\rangle ^{\otimes k})$ (equaling $\left| n/2-k \right\rangle$ in  the previous notation).  Any
  symmetric state $| \psi \rangle $ can therefore be written as
  \begin{equation}
    \label{eq:21}
    \left| \psi \right\rangle = \sum_{k=0}^{n} c_k \left| n,k \right\rangle.
  \end{equation}

  We now trace out $n-s$ qubits to calculate the reduced density
  operator $\rho_s$ of $s$ qubits. Since we deal with symmetric
  states, the choice of the qubits to trace out has no influence on
  the result.  It is convenient to use the general formula for
  splitting up a Dicke state into two sub-ensembles. It reads
  \begin{equation}
    \label{eq:23}
    \left| n,k \right\rangle = \sum_{l=0}^s \sqrt{\frac{\binom{s}{l}\binom{n-s}{k-l}}{\binom{n}{k}}} \left| s,l \right\rangle \otimes \left| n-s,k-l \right\rangle.
  \end{equation}
  We use this equation to express $\rho_s$ as
  \begin{equation}
    \label{eq:22}
    \rho_s = \mathrm{Tr}_{s+1,\dots,n} \left| \psi \right\rangle\!\left\langle \psi\right| 
    = \sum_{k,k^{\prime}} c_k^{*}c_{k^{\prime}}
 \sum_{l,l^{\prime}}\sqrt{\frac{\binom{s}{l}\binom{s}{l^{\prime}}\binom{n-s}{k-l}\binom{n-s}{k^{\prime}-l^{\prime}}}{\binom{n}{k}\binom{n}{k^{\prime}}}}
      \left| s,l \right\rangle\!\left\langle s,l^{\prime}\right|
      \delta_{k-l,k^{\prime}-l^{\prime}}.
  \end{equation}
  By summing over $k^{\prime}$ and shifting one summation index $k
  \rightarrow k+l$, this simplifies to
  \begin{equation}
    \label{eq:24}
    \rho_s = \sum_{l,l^{\prime}=0}^{s}\sum_{k=0}^{n-s} c^{*}_{k+l}c_{k+l^{\prime}}\binom{n-s}{k}\sqrt{\frac{\binom{s}{l}\binom{s}{l^{\prime}}}{\binom{n}{k+l}\binom{n}{k+l^{\prime}}}}\left| s,l \right\rangle\!\left\langle s,l^{\prime}\right|.
  \end{equation}

  \subsection{One-axis twisted spin-squeezed states}
  \label{sec:general-formalism}

  Equation (\ref{eq:24}) is now evaluated for $| \mathrm{S}(\mu) \rangle $ from Eq.~(\ref{eq:14}). Since we are only interested in how entangled $\rho_s$ is, $\nu$ is set to zero in the following. Then one has $c_k = 2^{-n/2}\sqrt{\binom{n}{k}} \exp[-i \frac{1}{2}\mu (k-n/2)^2  ]$. Plugging this into Eq.~(\ref{eq:24}), one can easily sum over $k$. Then one has 
\begin{equation}
\label{eq:26}
\rho_s = \sum_{l,l^{\prime}=0}^s \frac{\sqrt{\binom{s}{l}\binom{s}{l^{\prime}}}}{2^s} e^{-i\mu[l(s-l) - l^{\prime}(s-l^{\prime})]/2} \cos^{n-s}[\frac{1}{2}\mu(l-l^{\prime})]\left| s,l \right\rangle\!\left\langle s,l^{\prime}\right|.
\end{equation}
For $\mu = 0$, $\rho_s$ is separable as it equals $\left| + \right\rangle\!\left\langle +\right| ^{\otimes s} = 2^{-s}\sum_{l,l^{\prime}} \sqrt{\binom{s}{l}\binom{s}{l^{\prime}}}\left| s,l \right\rangle\!\left\langle s,l^{\prime}\right|$. For $\mu>0$, the state is entangled. However, for fixed $\mu$ and $s$, the contribution from the cosine is exponentially suppressed in $n$. However, the state $2^{-s}\sum_{l} \binom{s}{l}\left| s,l \right\rangle\!\left\langle s,l\right|$ is also separable, as it can be written as a convex sum of separable states. (Expressed differently, this state results from a spin-coherent state after complete dephasing.) Therefore, we see that by increasing $n$ and keeping the other parameters fixed, $\rho_s$ is exponentially close to a fully separable state.

\subsection{Dicke states}
\label{sec:dicke-states}

Dicke states $| m \rangle $ are such that for $k = n/2 - m$ we have $c_k = 1$ and zero for the other coefficients. Using Eq.~(\ref{eq:21}), we simply find that 
\begin{equation}
\label{eq:27}
\rho_s = \sum_{l = 0}^s \frac{\binom{s}{l}\binom{n-s}{k-l}}{\binom{n}{k}} \left| s,l \right\rangle\!\left\langle s,l\right|.
\end{equation}
In the limit of large $n$ and fixed $s$ and $k$, $\rho_s$ approaches the state $\sum_l \binom{s}{l} (k/n)^l (1-k/n)^{s-l} \left| s,l \right\rangle\!\left\langle s,l\right|$, which is a binomially distributed incoherent sum of Dicke states. As for the spin-squeezed state, one can easily show that this state is a dephased coherent state and therefore separable. In contrast to spin-squeezed states, however, the decay of entanglement is slower. We show this for the example of bipartite entanglement. The reduced two-qubit state reads
\begin{equation}
\label{eq:28}
\rho_2 = \frac{1}{n(n-1)}\left[ (n-k)(n-k-1) \left| 2,0 \right\rangle\!\left\langle 2,0\right| + 2k(n-k) \left| 2,1 \right\rangle\!\left\langle 2,1\right| + k(k-1)\left| 2,2 \right\rangle\!\left\langle 2,2\right|  \right].
\end{equation}
The entanglement of $\rho_2$ in terms of negativity $N$ \cite{Vidal_Computable_2002} is easy to compute analytically. For $k=1$ (i.e., the W state), it reads $N = (2-n+\sqrt{8-4 n+n^2})/(2n) = O(n^{-2})$; for $k = n/2$ (which is the most nonclassical state among the Dicke states), one finds $N = 1/(2n-2) = O(n^{-1})$. Generally, a good approximation for large $n$ is given by 
\begin{equation}
\label{eq:29}
N \approx \frac{k(n-k)}{n[n^2- 2k(k-n)]}.
\end{equation}
In conclusion, the entanglement of reduced states of Dicke states decays with a power law, which is in contrast to oversqueezed squeezed states, whose reduced states are exponentially close to separable states.

\end{widetext}

\bibliographystyle{apsrev4-1}
\bibliography{ImprovedHeisenberg}
 
\end{document}